\documentclass{elsart}

\def\deg{$^{\circ}$\,}

\def\ala{\alpha_{\rm A}}
\def\dea{\delta_{\rm A}}
\def\alb{\alpha_{\rm B}}
\def\deb{\delta_{\rm B}}

\def\cda{\cos \dea}
\def\csqda{\cos^2 \dea}

\def\sda{\sin \dea}

\def\cdb{\cos \deb}
\def\csqdb{\cos^2 \deb}

\def\sdb{\sin \deb}

\def\cabaa{\cos (\alb - \ala)}
\def\Arc{{\mathcal{A}}}

\def\sigarc{\sigma_{\Arc}}

\def\bea{\begin{eqnarray*}}
\def\eea{\end{eqnarray*}}
\def\beqna{\begin{eqnarray}}
\def\eeqna{\end{eqnarray}}
\def\btab{\begin{tabular}}
\def\etab{\end{tabular}}

\newcommand{\devpar}[2]{\frac{\partial #1}{\partial #2}}
 
\usepackage{natbib}

\usepackage{graphicx}
 
\begin{document}

\runauthor{P\'erez-Torres, Marcaide,  Guirado, and Ros}

\begin{frontmatter} 
\title{Differential Astrometry over 15$^\circ$} 
\author[DAA]{M.A. P\'erez-Torres}
\author[DAA]{J.M. Marcaide}
\author[DAA]{J.C. Guirado}  
\author[MPI]{E. Ros}  


\address[DAA]{Depto. de Astronom\'{\i}a,  
Universitat de Val\`encia, 
E-46100 Burjassot, Spain } 
\address[MPI]{Max-Planck-Institut f\"ur Radioastronomie, Auf 
dem H\"ugel 69, Bonn, Germany}

\begin{abstract} 
  
We observed the pair of radio sources 1150+812 and 1803+784 in
November 1993 with a VLBI array, 
simultaneously recording at 8.4 and 2.3 GHz.
We determined the angular separation between the two sources
with submilliarcsecond accuracy
by using differential techniques.
This result demonstrates the feasibility of high precision
differential astrometry for radio sources separated
in the sky by almost 15$^\circ$, and opens the avenue
to its application to larger samples of radio sources.

\end{abstract} 



\end{frontmatter} 

\section{Introduction}
\label{intro} 

Very Long Baseline Interferometry (VLBI) 
is a powerful astrometric technique.                   
Centimeter-wavelength VLBI group-delay astrometry of 
extragalactic radio sources 
routinely provides precisions at 
the milliarcsecond (mas) level, thus allowing a 
celestial reference frame to 
be built with milliarcsecond accuracy (Ma et al.  1990). 
One step further in high precision astrometry is phase-delay differential 
astrometry, in which the differences between the phase delays 
of two radio sources
are used to determine their relative separation. 
For close source pairs, this technique yields accuracies 
of a few microarcseconds ($\mu as$), i.e. the case 
of the double quasar 1038+528 A and B  (Marcaide \& Shapiro, 1983), 
whose components are separated by less than an arcminute.
For sources separated up to a few degrees, accuracies of about
0.1--0.3 mas are achieved (Guirado et al. 1995, 
Lara et al. 1996, Ros et al. 1999a). 
The basic idea underlying differential astrometry is that 
the differenced phase-delays from the source pair will be 
largely free of unmodelled systematic effects.
If the sources are very close, cancellation of systematic effects is
almost complete, whereas if 
the sources lie at increasingly angular distances, the cancellation 
is less perfect, and the determination of 
their angular separation is correspondingly less accurate.

In this contribution, we address the question of the applicability 
of the differential astrometric technique to a 
pair of strong radio sources, 1150+812 and 1803+784,  
separated by almost 15$^\circ$.
We demonstrate the feasibility of the technique for 
this pair of radio sources, and measure their relative 
separation with submilliarcsecond accuracy.
In addition, we show that the standard errors are dominated 
by uncertainties in the coordinates of the reference source.

\section{Observations}
\label{obs} 

We observed the radio sources 1150+812 (QSO) 
and 1803+784 (BL Lac)
on November 1993 during 12 hours,
simultaneously recording at 8.4\,GHz and 2.3 GHz.  
We used the following antennas: 
Effelsberg, Medicina, Onsala, Fort Davis, Hancock, 
Los Alamos, North Liberty, Owens Valley, and the Phased VLA.
The data were correlated at the Max-Planck-Institut 
f\"ur Radio\-astro\-no\-mie, Bonn, Germany.
We used an observing cycle of 7 min, consisting of 
2 min  observing 1150+812,
1.5 min of antenna slew, 2 min observing 1803+784, 
and 1.5 min slew back to 1150+812.
Since both sources were strong ($>$ 1 Jy) at the epoch of observation,
we have detected the radio sources for all of the baselines, with
high signal-to-noise-ratio (SNR) at both frequencies. 

After correlation, we exported the data 
and  constructed the interferometric visibilities 
using the calibration information provided by the 
staff at the observing antennas.
We used the software package DIFMAP (Shepherd et al., 1995) 
to obtain images of both sources at each frequency.
The images (Figs.\ 1 and 2) clearly show the typical core-jet 
structure for this kind of radio sources, the 
brightest spot being usually associated with the 
flat-spectrum core, and the others 
with steeper spectrum components of a jet. 
1150+812 displays a southeastward jet, while 
1803+784 displays a westward jet.

\begin{figure}[htb]
\vspace{-5.0cm}
\centering
\includegraphics[scale=0.65,angle=0]{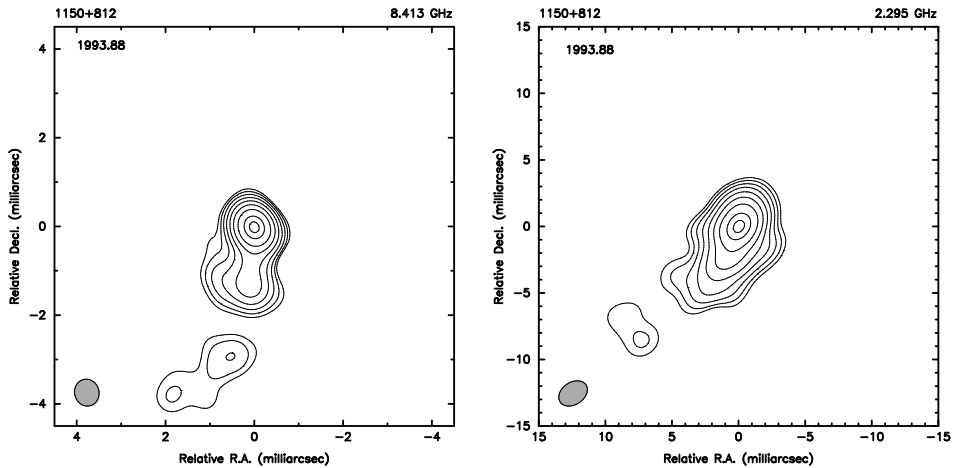}  
\vspace{-5.5cm}
\caption{
Hybrid maps of 1150+812 at 8.4 and 2.3 GHz.
Contours are 0.5,1,2,4,8,16,32,64 and 90\% of the peak of
brightness, 0.90 Jy/beam and  0.74 Jy/beam for 8.4 and 2.3 GHz, respectively. 
The sizes of the natural restoring beams are 0.61 $\times$ 0.55 mas 
(PA=15$^\circ$) at 8.4 GHz, and 2.33 $\times$ 1.67 mas (PA=--56$^\circ$) 
at 2.3 GHz.
East is left, and North is up. Note that scales are different in both maps. 
}
\label{fig1}
\end{figure}

\section{Astrometric Data Reduction}

The goal of the data reduction is to obtain a data set 
of differenced phase-delays on which a
weighted-least-squares algorithm can be applied to determine the 
relative position of the sources.
To obtain this data set, we went 
through several steps, namely, 
phase connection (or correction of the 
$2\pi$ ambiguities of the phases; Shapiro et al., 1979), 
structure correction, 
and removal of opacity and propagation medium effects.
In the astrometric process, we used data from the antennas 
with best performance 
(Effelsberg, Medicina, Fort Davis, Los Alamos, North 
Liberty, and Owens Valley). 

We used our hybrid maps to correct for the source structure
contribution to the phase delays, using as reference point the 
peak of the brightness distribution of each map. 
Most of the propagation medium effects are caused by 
the troposphere and the ionosphere. 
We accurately modelled the tropospheric behaviour 
from detailed information of the weather parameters at 
each of the antenna sites. 
The ionosphere is a refractive plasma that introduces a 
dispersive contribution to the phase delays, 
which scales as $\nu^{-2}$.
Thus, we used our dual-frequency data to estimate and remove 
such contribution. 
For it,  the reference points on the 2.3 GHz maps should 
correspond to those chosen on the 8.4 GHz maps. We took into account 
opacity effects to obtain a reliable map registration
(Guirado et al., 1998). 

In this way, we obtained a set of (undifferenced) phase delays, 
free of structure and propagation medium effects. 
However, the existence of unmodelled systematic effects 
was evident from a visual inspection of the data. 
We formed
a set of differenced phase delays by subtracting the delay of 
each observation of 1150+812 from the delay of the previous 
observation of 1803+784. 
Although the sources are separated by 15\deg , this procedure
largely cancelled out systematic effects still present in the undifferenced 
phase delays.
        
\begin{figure}[tbh]
\vspace{-5.0cm}
\centering
\includegraphics[scale=0.65,angle=0]{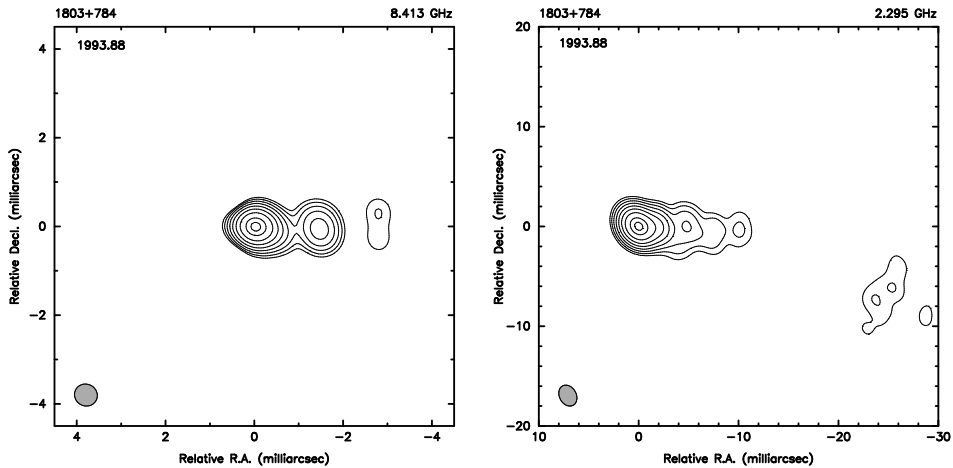}  
\vspace{-5.5cm}
\caption{
Hybrid maps of 1803+784 at 8.4 and 2.3 GHz.
Contours are 0.5,1,2,4,8,16,32,64, and 90\% of the peak of
brightness, 1.48 Jy/beam and 1.35 Jy/beam for 8.4 and 2.3 GHz, 
respectively. The size of the natural restoring beam is 
0.52 $\times$ 0.49 mas (PA=61$^\circ$) at 8.4 GHz, and  
2.21 $\times$ 1.68 mas (PA=29$^\circ$) at 2.3 GHz.
East is left, and North is up. Note that scales are different in both maps. 
}
\label{fig2}
\vspace{0.5cm}
\end{figure}

\section{Relative position of the radio sources}
\label{sec:relpos}

Our estimates of the relative positions of 
1150+812 and 1803+784, in right ascension and declination,  
are based on a weighted-least-squares
analysis of the differenced phase delays. 
We also included the (undifferenced) phase delays for 1803+784 
in the analysis to estimate the behaviour of the station clocks. 
The standard deviations of the differenced phase delays, 
and of the 1803+784 phase delays were scaled separately so that, 
for each baseline, the root-mean-square of the postfit residuals was unity.  
The difference between our estimates  
and those given by IERS (1996) are shown in Table 1.  

\vspace{0.2cm}

\begin{table}[h]
\begin{center}
\caption{Estimates of the angular separation of 
1150+812 and 1803+784}
\vspace{0.5cm}
\begin{tabular}{ccrc}
\hline
$\Delta\alpha - 6^h 7^m 33\rlap{.}^s 18473$ &   
                   = & $ -0\rlap{.}^s 00015 \,  \pm$  $0\rlap{.}^s 00012$  \\
$\Delta\delta + 2^\circ 30' 25\rlap{.}'' 13601$ & 
                   = &  $ 0\rlap{.}'' 00040 \,  \pm$  $0\rlap{.}'' 00030$ \\
\hline
\end{tabular}
\end{center}
\end{table}

The error bars include 
formal uncertainties from a covariance analysis, and residual errors of 
the geometry of the interferometric array, source position, 
source opacity, and atmosphere.
The largest contributor
to the error budget in the relative separation 
of the sources is the uncertainty in the a priori position of the 
reference radio source, 1803+784 (IERS, 1996).

\section{The separation arclength}

Since the angular separation between the two sources is very large,
the use of the arclength, $\Arc$,  between the sources
could be a better estimate  of their relative separation than
$\Delta\alpha$ and  $\Delta\delta$, since $\Arc$ is invariant to
rotations and, therefore, independent of the reference system
where the source coordinates are defined.
The arclength between the two radio sources, compared to the
arclength as given by IERS is  

\begin{eqnarray*}
\Arc - 14^\circ 50' 21''.14759 =  0''.00040 \pm \, 0''.00019 
\label{eq:arcl}
\end{eqnarray*}

The uncertainty of the arclength estimate is smaller than 
for $\Delta\alpha$ and $\Delta\delta$, as expected.
However, the special geometry of the radio sources (see below) 
prevents it from being even smaller. 
To better clarify this point, we develop 
the expression for the uncertainty in the determination
of the arclength, $\sigarc$, using an error propagation 
equation (Bevington \& Robinson, 1992), from

\beqna
\Arc(\ala,\alb,\dea,\deb) & = & \arccos[ \cda \cdb \cabaa 
                         + \sda \sdb] \nonumber \\
\sigma_\Arc^2 &  = &  \sum_{x_i} \sigma_{x_i}^2 
                      \left(\devpar{\Arc}{x_i}\right)^2 + 
                \sum_{x_i \neq x_j} 2\sigma_{x_i x_j}^2  
                \left(\devpar{\Arc}{x_i}\right) 
                \left(\devpar{\Arc}{x_j}\right) 
\label{eq:sigma1}
\eeqna

where $\sigma_{x_i x_j}^2 \approx \sigma_{x_i} \sigma_{x_j} \rho (x_i,x_j) $ 
is the  covariance of coordinates $x_i$  and $x_j$
(each one taking the values $\ala , \alb, \dea , \deb $), 
$\sigma_{x_i}$ are the a priori uncertainties of the source 
coordinates,  
and $\rho (x_i,x_j)$ are the correlation coefficients. 
The North Celestial Pole, and the sources  
1150+812 (source A) and 1803+784 (source B, taken as the reference source),  
form a triangle of $\sim 90^\circ$ at the    
North Celestial Pole.
The geometry is such that $\cabaa \ll 1; \cda, \cdb \ll 1$. 
We can then write Eq.~\ref{eq:sigma1} approximately as 

\begin{eqnarray*}
(\sigma_\Arc / \sigma_{\deb})  & \approx 
            \sqrt\frac{K^2 \csqda + \csqdb}{\csqda + \csqdb}
\label{eq:sigma2}
\end{eqnarray*}

where $ K = \sigma_{\dea} / \sigma_{\deb} = 1.4$. This gives  
$\sigma_\Arc \approx 1.18 \, \sigma_{\deb} \approx 0.17$ mas.  
(Note that K=1 would imply $\sigma_\Arc \approx \sigma_{\deb}$). 
This result shows that $\sim 89$ \% of the overall standard 
deviation of the arclength is produced solely by the contribution
of the a priori standard deviations of the source coordinates. 
We emphasize that this is a consequence of the location of the 
radio sources in the celestial sphere, and dominate by far the 
standard deviation of our astrometric determination.

\section{Results and Conclusions}

We observed the pair of strong radio sources 1150+812 and 1803+784 
with an interferometric VLBI array on November 1993. 
The antennas simultaneously recorded data at 8.4 and 2.3 GHz, 
which allowed us to remove ionospheric contributions to 
the delays. 
We also estimated the contributions to the delays
due to the troposphere, source structures, and source opacities.
From geometric considerations, we infer
that the main contribution to the uncertainty in
the determination of the relative angular separation 
comes from the uncertainty in the a priori position of the 
reference source. We show that this uncertainty is further increased by the 
particular geometry of the radio source pair.

We determined with submilliarcsecond accuracy 
the angular separation of the pair of radio sources, 
separated by almost 15$^\circ$, and 
thus demonstrated the feasibility of using phase delays
from dual-frequency VLBI measurements for very large angular 
separations. 
This accomplishment  opens the applicability of the technique to a 
large sample of radio sources (Ros et al., 1999b).
A final goal of building a quasi inertial celestial reference 
frame based on stationary radio source cores, 
with submilliarcsecond accuracy, appears within reach.

\end{document}